\documentclass[namedreferences]{kluwer}
\usepackage{epsfig}

\def\etal{{\it et al.~}}

\begin{document}
\begin{article}
\begin{opening}
\title{YOUNG POPULATIONS IN THE NUCLEI OF BARRED GALAXIES}
\author {H. B. \surname{Ann}\email{hbann@cosmos.es.pusan.ac.kr}}
\institute{Dept. of Earth Science, Pusan National University}
\runningtitle{YOUNG POPULATIONS IN BARRED GALAXIES}
\runningauthor{ANN}
\begin{ao}
Department of Earth Science,
Pusan National University,
Pusan, 609-735, Korea
e-mail: hbann@cosmos.es.pusan.ac.kr
Fax: +82 51 5137495
\end{ao}
\begin{abstract}
We have conducted a $UBVRI$ and $H_{\alpha}$ CCD photometry of 5 barred
galaxies, NGC2523, NGC2950, NGC3412, NGC3945 and NGC5383,
along with SPH simulations, in order to
understand the origin of young stellar populations in the nuclei of
barred galaxies.  The $H_{\alpha}$ emissions which are thought to be emitted
by young stellar populations are either absent or strongly concentrated on the
nuclei of early type galaxies (NGC2950, NGC3412, NGC3945),
while they are observed in the nuclei and circumnuclear regions of
intermediate type galaxies with strong bars (NGC2523, NGC5383).
SPH simulations of realistic mass models for these galaxies show that 
some disk material can be driven into the nuclear regions by
strong bar potentials.  
This implies that the young stellar populations in the circumnuclear regions 
of barred galaxies can be formed out of such gas.
The existence of the nuclear dust lanes is an indication of an on-going gas 
inflow and extremely young stellar populations in these galaxies,
because the nuclear dust lanes such as
those in NGC5383 are not long-lasting features according to our simulations. 
\end{abstract}
\keywords{galaxies, photometry, population, SPH simulations}
\end{opening}
\section{INTRODUCTION}
Barred galaxies are characterized by the prominent bars across 
the nuclei of disk galaxies. 
Although bulges of disk galaxies are thought to consist of old
stellar populations formed in the early phase of galaxy evolution,
there are some disk galaxies, especially early type barred galaxies,
which have young stellar populations in their nuclei (Benedict \etal 1992).
The association of bars with nuclear young stellar populations which are
delineated as "hot spots" was recognized first 
by Sersic \& Pastoriza (1965, 1967).
Most of the galaxies with young stars in their nuclei show twisted isophotes,
indicating triaxial structures (Stark 1977; Wozniak \etal 1995).
Hence Wozniak \etal (1995) classified the nuclear morphology
of galaxies according to the characteristics of isophotal twists.

Recent hydrodynamical simulations have shown that strong bars are very
effective in transporting the disk material into the inner regions of
galaxies, and so the young stellar populations can form from the gas infalled
into the nuclear region (Friedli \& Benz 1993; Ann \& Kwon 1996).
Thus a detailed analysis of star formation rates and the distribution
of the young stellar populations in observed galaxies may provide
observational constraints for the gas inflow model driven by a bar potential.
The purpose of the present photometry is to draw some insights into the physical
conditions that allow young stellar populations in the nuclei of galaxies.
Toward this end, we have conducted $UBVRI$ and $H_{\alpha}$ photometry
of five bright barred galaxies, and then derived realistic mass models
from the observed surface brightness.  We have also carried out
an extensive suite of hydrodynamic simulations of barred disk galaxies
by adopting the derived mass models.

\section{OBSERVATIONS AND REDUCTION}
The observations of the program galaxies were made on the two photometric
nights during the observing run in March 1997 with a
Tex 1024 CCD ($1024 \times 1024$, $24^2$ $\mu\rm{m^{2}}$ pixels) attached at
the Cassegrain focus of the 1.8 m Doyak Telescope of Bohyunsan Optical
Astronomy Observatory (BOAO) in Korea. 
The gain and readout noise of the CCD were $3.5 e^{-1}$/ADU and $6.4e^{-1}$,
respectively.  The images were recorded with $2 \times 2$ on-chip binning
which gives the image scale of 0.$^{''}$68/pixel.
The images were obtained with Johnson-Cousin {\it UBVRI} and {\it $H_\alpha$}
filters. Flat field exposures were made on the twilight sky. 
Several standard stars of Landolt (1992) and Massey \etal (1988) were observed
during the nights for the absolute calibration of the sky brightness and
$H_{\alpha}$ flux.  The seeing was about 2$^{''}$ throughout the nights.

The basic reductions were carried out using the CCDRED package within IRAF.
The flat fielded frames of galaxy images were subtracted and divided
by sky frames which were obtained by fitting the sky regions surrounding the
galaxy images.  We applied a variable width Gaussian smoothing to increase the
signal-to-noise ratios of the observed data in the outer part of the galaxy.
The $H_{\alpha}$ flux calibrations were made by subtracting the scaled
$R$-band images from the $H_{\alpha}$ images.  The scale factors were
determined by the stars and the regions devoid of emissions in the galaxy.

\begin{figure}
\centerline{
}
\caption{$V$-band isophotal maps of NGC~2523, NGC~2950, NGC~3412, NGC~3945
and NGC~5383.  All of the galaxies except NGC~2523 seem to have triaxial bulges
which are aligned perpendicular to the primary bars.
The sizes of the maps are $2.^{\prime}23 \times 2.^{\prime}23$.
North is up and east to the left.}
\end{figure}
\section{NUCLEAR MORPHOLOGY AND YOUNG POPULATIONS}
The presence of young stellar populations in the nuclear region of 
observed galaxies can be inferred from the nuclear structures such as nuclear 
ring, nuclear spiral arms and nuclear dust lanes.  Usually these structures
are associated with HII regions of newly formed massive stars. These nuclear
structures distort the shape of the isophotes in the nuclear region and make
the appearance of the bulges triaxial (Shaw \etal 1995; Wozniac \etal 1995).

Fig.~1 displays the $V$-band isophotal maps of the five barred galaxies.
We can see pronounced isophotal
twists in the nuclear regions, which suggest
the bulges of the three SB0 galaxies (NGC~2950, NGC~3412 and NGC~3945) are
triaxial. 
But the fact that there is no nuclear structure such as those mentioned above
implies no young stellar population in their bulges. 
However, NGC~5383 shows very complicated
structures which resemble nuclear spiral arms and nuclear ring.
Detailed analysis of the nuclear morphology of NGC~5383 (Ann and Kim 1998)
showed that the spiral patterns are made by the sharply curved nuclear
dust lanes obscuring the nuclear ring.  Since dust lanes are located
in the high density regions where stars can form, we expect young stellar
populations in the nuclear region of NGC~5383.

The strong $H_{\alpha}$ emission can provide 
a more direct evidence for the presence of the young stellar population. 
As shown in Fig.~2, there are strong $H_{\alpha}$ emissions 
in the nuclear regions of two intermediate type galaxies NGC~2523 and NGC~5383. 
Most of the $H_{\alpha}$ emissions in these galaxies come from the 
circumnuclear regions which are similar to the nuclear rings such as those in 
NGC~4314 (Benedict \etal 1992; Ann \& Kwon 1996) and 
NGC~3351 and NGC~4303 (Colina \etal 1997).  However, the shape of the nuclear
ring of NGC~2523 is very different from that of NGC~5383 due to the 
different bulge morphologies of two galaxies. 
There are virtually no $H_{\alpha}$ 
emissions in the bulges of SB0 galaxies NGC~2950 and NGC~3412,  
while the $H_{\alpha}$ emissions are confined to the nucleus in NGC3945. 
It is evident from the surface brightness profiles in Fig.~2 that the 
distribution of the old stellar populations, which contribute most of the 
luminosities 
in the $UBVRI$ pass bands, is quite different from that of the young stellar
populations inferred from the $H_{\alpha}$ profiles.  The contribution 
of the young populations to the surface brightness profiles is almost
negligible in the longer wavelengths.  

\begin{figure}
\centerline{
}
\caption{Surface brightness profiles of five barred galaxies.
There seem to be nuclear rings in NGC~2523 and NGC~5383 which show
strong $H_{\alpha}$ emissions.  In NGC~3945 the strength of $H_{\alpha}$
emissions is quite high but they are confined to the nucleus.  There
is virtually no $H_{\alpha}$ emission in NGC~2950 and NGC~3412.}
\end{figure}
\section{SPH SIMULATIONS}
We have conducted SPH simulations to understand the origin of the young
stellar populations in the nuclear region of the barred galaxies.
The basic numerical methods employed in our SPH simulations are the same as
those of Ann \& Kwon (1996) and Lee \etal (1999).  We assumed the model 
galaxies consisted of four components: bulge, disk, primary bar,
and nuclear bar.  We adopted the exponential disk of Freeman (1970) for the 
stellar disk component and the Plummer spherical potential for the bulge.
Because our simulation was restricted to the two-dimensional disk,
we used the bi-axial bar potential of Long \& Murali (1992),
which was uniform in space and constant in time.  We assumed that
the gaseous disk was isothermal at $T=10^4$ K.
The number of SPH particles was about 10000. 

We have considered the following two models that can reproduce the observed 
morphologies of NGC~5383 (Model~A) and NGC~2523 (Model~B). 
The mass fractions in different mass components were 
derived from the decomposition of the observed surface brightness profiles 
into bulge, disk and bar. 
\begin{itemize}
\item[]{\it Model A}
 
$M_{d}:M_{bg}:M_{bar}=3.6:3.4:2.7$, a=0.5, a/b=4, $\Omega =2$,\par
$R_d=0.4$,~b=0.1,$M_{gas}$=0.02, $M_{sbar}$=0.01
\item[]{\it Model B}
  
$M_{d}:M_{bg}:M_{bar}=4.0:2.8:3.0$, a=0.4, a/b=4, $\Omega =2.5$,\par
$R_d=0.3$,~b=0.09, $M_{gas}$=0.02, $M_{sbar}$=0.001
\end{itemize}
\noindent{where a is bar major axis lengh, a/b is axial ratio, $R_d$ is disk scale
length and $M_{sbar}$ is the mass fraction of the nuclear bar.
In our simulations, the physical quantities are expressed as dimensionless
variables with the following parameters;}
\begin{itemize}
\item[]{}
$M_{tot} =2 \times 10^{11}M_\odot$, $R_{sc} = 15 kpc$,
$\tau_{sc} = {1 \over \sqrt{G\bar\rho}} = 1.2 \times 10^8 yrs$,\par
$\Omega_{sc} = 16.4 km/sec/kpc$
\end{itemize}

Fig.~3 shows The distribution of the SPH particles
and the mass fraction of the gas within 4.5 kpc from the center of the 
model galaxies.  The bar
pattern speeds are chosen to be low enough to ensure the ILRs for the model
galaxies. One can see in the middle panel of Fig.~3 that the distributions 
of SPH particles resemble the nuclear morphology of NGC~5383 and NGC~2523.
Moreover, the dust lanes in the bar of NGC~5383 are also seen in the gas
distribution of Model~A.  This means that our models quite well represent
the inner dynamics of the real galaxies.
Both of our models lead to a significant gas infall toward the nuclear 
regions, which enables nuclear star formation there.
The mass infall rate calculated from numerical simulations
is order of $\sim 0.5 \times  M_\odot yr^{-1}$ for both models with a
strong primary bar ({\it i.e.} $M_{bar}/M_{tot} \sim 0.3$ and a/b=4).
This is consistent with the star formation rate inferred from the
$H_{\alpha}$ emissions of the observed galaxies.

\begin{figure}


\caption{The distribution of test particles and the mass fraction of
infalled gas within inner regions (from top to bottom). Model~A has a
more massive bulge compared to Model~B.}
\end{figure}

\section{SUMMARY AND DISCUSSION }
Three SBO galaxies (NGC~2950, NGC~3412, NGC~3945) seem to have virtually
no young stellar populations in the bulges except for the nuclei where
some $H_{\alpha}$ emissions are detected.  The $H_{\alpha}$ emissions from
the nucleus of NGC~3945 is much stronger than those from the nuclei of
NGC~2950 and NGC~3412 which have weaker bars than that of NGC~3945.
Two intermediate-type barred galaxies, NGC~2523 and NGC~5383, with  
strong bars show strong $H_{\alpha}$ emissions in the nuclear
regions, indicating a significant number of young stellar populations
there.

However, the elongated bulges of the three SBO galaxies are thought to 
result from the early secular evolution driven by the bars, that is, 
transport of disk material into the nuclear regions.  
The absence of the young stellar populations in their nuclear regions implies 
lack of the gas in the disks of these galaxies.  
There might have been some mechanisms
which blew out the remnant gas from the galaxies after triaxial bulges
were formed by the secular evolution driven by the bars. 

Numerical hydrodynamic simulations of barred disk galaxies have
shown that a bar potential can drive a spiral structure and non-
axisymmetric flow motions along with shocks.
The shocked gas loses its angular momentum and kinetic energy and
spirals in toward the galactic center.  Our SPH simulations 
based on realistic mass models show that strong bars with axial ratios
larger than 3 are very effective in transporting the disk material
into the nuclear regions of galaxies.  

The nuclear rings can be made of young stars formed from the infalled
gas with their morphology depending on the mass fractions in the bulge
and the disk.  The nuclear rings formed in galaxies with large 
bulge-to-disk ratio are aligned perpendicular to the primary bar 
while those with small bulge-to-disk ratio
are aligned with the bar axis.  

\begin{acknowledgements}
We thank Hyesung Kang for critical reading of the manuscript.
This work was supported in part by the Basic Science
Research Institute Program, Ministry of Education, BSRI-98-5411, and
in part by Korea Research Foundation non-directed research grant No.
1997-001-D00163.
\end{acknowledgements}
{}
\end{article}
\end{document}